\documentclass[12pt]{article}
\begin{document}

\newcommand{\re}{\mathop{\mathrm{Re}}}

\newcommand{\be}{\begin{equation}}
\newcommand{\ee}{\end{equation}}
\newcommand{\bea}{\begin{eqnarray}}
\newcommand{\eea}{\end{eqnarray}}

\begin{center}

{\large \bf An interesting property of the Friedman universes}

\vspace{2.cm}

{Janusz Garecki\footnote{E-mail address:
garecki@wmf.univ.szczecin.pl}}\\
{\it Institute of Mathematics, University of Szczecin, Wielkopolska 15,
          70-451 Szczecin, Poland}

\end{center}

\date{\today}
\vspace{0.3cm}
\begin{center}
This work is a slightly amended lecture delivered at {\bf Hypercomplex Seminar 2012}
10-15 July 2012, at B\c edlewo, Poland
\end{center}
\vspace{0.3cm}

\input epsf

\begin{abstract}
We show in the paper that Friedman universes can be created from empty,
flat Minkowskian spacetime by using suitable conformal rescaling of the spacetime metric.
\end{abstract}




\newpage

\section{Friedman universes}

Einstein equations and Cosmological Principle lead us together to Friedman universes.
These universes give standard mathematical models of the real Universe.

Einstein equations
\begin{equation}
G_{ik} := R_{ik} - {1\over 2}g_{ik} R ={8\pi G\over c^4}
T_{ik}=:\beta T_{ik}
\end{equation}
form system of the ten, 2-nd order quasilinear partial
differential equations on ten unknown functions. Solving these
equations under given initial and boundary conditions one obtains
local geometry of the spacetime, i.e., $g_{ik}(x)\longrightarrow
\Gamma^i_{~kl}(x)\longrightarrow R^i_{~klm}(x)$  and local
distribution and motion of matter, i.e., $T_{ik}(x)$.

Here $G_{ik}$ is the so-called {\it Einstein tensor}, $T_{ik}$
is the {\it matter energy-momentum tensor} (the source of the gravitational field
which is represented by tensor $G_{ik}$), $c$
is the velocity of light in vacuum, and $G$ means Newtonian
gravitational constant; $g_{ik}(x)$ denote components of the metric
tensor, and $\Gamma^i_{~kl}(x), ~~R^i_{~klm}(x))$ are the Levi-Civita
connection and Riemannian curvature components respectively. $R_{ik}$
mean components Ricci tensor and $R$ is the so-called curvature
scalar (See, eg., \cite{Sok}). All Latin indices take values
$0,1,2,3$.

The matter tensor $T_{ik}(x)$ consists of $g_{ik}, ~u^i,
~p,~\rho$, where $u^i,~p, ~\rho$ denote 4-velocity, pressure and
density of matter respectively.

Cosmological Principle says that in the largest scale the real
Universe is homogeneous and isotropic\footnote{We are modelling {\it cosmological substrat}
by using an ideal (or perfect) fluid.}.

In the following we will use {\it geometrized units} in which $G=c
=1$.
Friedman universes are cosmological solutions to the
Einstein equations constrained by Cosmological Principle and they
are foundation of the relativistic cosmology \cite{Sok,And}.

The line element $ds^2 = g_{ik}(x) dx^i dx^k$ for these universes,
called {\it Friedman-Lemaitre'-Robertson-Walker} line element, in
the {\it comoving coordinates} $x^0 = t, ~x^1 = \chi,~~x^2 =\vartheta,~~x^3
=\varphi$, reads
\begin{equation}
ds^2 = dt^2 -R^2(t)\bigl[d\chi^2 + S^2(\chi)\bigl(d\vartheta^2 +
sin^2\vartheta d\varphi^2\bigr)\bigr],
\end{equation}
where
\begin{eqnarray}
S(\chi) &=& sin\chi, ~~if ~~k = 1\nonumber\\
S{\chi}& =& \chi, ~~if~~ k = 0\nonumber\\
S(\chi)&=& sh\chi, ~~if~~ k =(-)1.
\end{eqnarray}

$t$ is the {\it cosmic time}, i.e., the proper time for {\it isotropic observers},
which are at rest in the coordinates $(t,\chi,\vartheta,\varphi)$.

An isotropic observer $O$ represents center of mass of a
cluster of galaxies in real Universe. $R(t)$ is the so-called {\it
scale factor} (it scales spatial distances) and $ k = 0,\pm 1$ means the {\it normalized
curvature} (curvature index) of the spatial sections $x^0 = t = const$.

If $k=1$, then we have closed (spherical or elliptical) spatial sections, if $k=0$ the
geometry of the spatial section is flat, and if $k=(-)1$, then the
geometry of spatial sections is hyperbolic.

Usually one chooses the moment $t = 0$ of the cosmic time $t$ when
$R= 0$, i.e., usually one has $R(0) = 0$.

Einstein equations with perfect fluid (incompressible fluid, without any viscosity and not
conducting heat) as source \footnote{A particle of this fluid
represents a cluster of galaxies in real Universe.} reduce, for
the FLRW line element (2)-(3) to the {\it Friedman equations}
\begin{equation}
{3{\dot R}^2\over R^2} + {3k\over R^2} = {\rho\over 2\beta},
\end{equation}
\begin{equation}
{{\dot R}^2\over R^2} + {{\ddot R}\over R} + {k\over R^2} = (-)
{p\over 2\beta}.
\end{equation}

Here $\beta = 8\pi$ (We use geometrized units), $\rho = \rho(t)$ means the rest density of
the fluid, and $p = p(\rho) = p(t)$ --- its pressure. ${\dot R} := {dR\over
dt}$, and ${\ddot R} := {d^2R\over dt^2}$.

Caloric equation $p=p(\rho)$ must be added to Friedman equations
(4)-(5) in order to get a determined system on the three unknown functions:
 $R = R(t), ~~\rho = \rho(t), ~~p =p(t)$.

Usually one considers solutions to the Friedman equations
(4)-(5) in the two extreme cases: $p=0$ (dust universes or matter dominant universes, in
short {\bf MDU}), and $p={\rho\over 3}$ (radiation dominant
universes, in short {\bf RDU}).

We will confine to solutions in these two extreme cases.
\begin{center}
Dust universes ({\bf MDU}) with $p=0$:
\end{center}
\begin{enumerate}
\item k =1 (closed universe). In this case we have parametric solution
\begin{eqnarray}
R&=& M\bigl(1 - \cos\eta\bigr),\nonumber\\
t &=& M\bigl(\eta-\sin\eta\bigr).
\end{eqnarray}
$0<\eta<2\pi$.
\item k=0 (flat universe). In this case
\begin{equation}
 R = \bigl({9M\over 2} t^2\bigr)^{1/3}, ~~0<t<\infty.
\end{equation}
\item k =(-)1 (open universe). In the case we also have parametric
solution
\begin{eqnarray}
R &=& M\bigl(\cosh\eta -1\bigr),\nonumber\\
t&=& M\bigl(\sinh\eta-\eta\bigr), ~~0<\eta<\infty.
\end{eqnarray}
\end{enumerate}

Here $\eta$ denotes a parameter and $M = (4/3)\pi R^3\rho$ is the
first integral of the Friedman equations. Physically $M$
is the mass contained inside of a ``sphere'' having volume $(4/3)\pi
R^3$.

\begin{center}
Radiation universes ({\bf RDU}) with $p = {\rho\over 3}$
\end{center}
\begin{enumerate}
\item k= 1(closed universe)
\begin{equation}
R = \sqrt{(2bt - t^2)}, ~~b:=\sqrt{{8\pi C\over 3}},~~0<t<2b,
\end{equation}
where $C = \rho R^4 = const>0$ is the first integral of the
Friedman equations in this case.
\item k =0 (flat universe)
\begin{equation}
R = \sqrt{2bt}, ~~0<t<\infty.
\end{equation}
\item k=(-)1 (open universe)
\begin{equation}
R =\sqrt{(2bt+t^2)}, ~~0<t<\infty.
\end{equation}

\end{enumerate}

Having $R= R(t)$ one can find $\rho(t)$ from the first integrals
and then  $p=p(t)$ from caloric equations.

It is believed that one of the {\bf MDU}  correctly describes
present stage of the Universe, and that one of the {\bf RDU}
correctly describes early Universe\footnote{The recent large--scale astronomical
observations seem favorize an accelerated flat model.}.

It is seen from (6)-(11) that the Friedman universes are singular
at least in one moment of the cosmic time $t$ (In this moment
$R=0$). These singularities are inevitable in classical general
relativity  (Theorems by Hawking and Penrose, and Senovilla \cite{Hell}); but
``quantized general relativity'' (loops quantum gravity) seems
remove these singularities (Ashtekar, Bojowald  and Lewandowski) \cite{Ban}.

\section{Conformal rescaling of metric  and conformally flat
spacetimes}

By {\it conformal rescaling}  of the metric $g$ we mean the following
transformation  (in established coordinates)
\begin{equation}
{\hat g}_{ab}(x) = \Omega^2(x) g_{ab}(x),
\end{equation}
where the {\it conformal factor} $\Omega(x)$ is dimensionless,
smooth and positive.

One can immediately get from (12) that
\begin{equation}
{\hat g}^{ab}(x) = \Omega^{(-)2}(x) g^{ab}(x),
\end{equation}
and, after some tedious calculations one can obtain other useful transformational formulas \cite{Gar1}.
For our future aims the following formulas will be needed
\begin{eqnarray}
{\hat R}^b_{~d}& =& \Omega^{(-)2} R^b_{~d}
+2\Omega^{(-)1}\bigl(\Omega^{(-)1}\bigr)_{;dc} g^{bc}\nonumber\cr
&-&{1\over 2}\Omega^{(-)4}\bigl(\Omega^2\bigr)_{;ac}
g^{ac}{}\delta^b_d,
\end{eqnarray}
\begin{equation}
{\hat R} = \Omega^{(-)2} R - 6\Omega^{(-)3} \Omega_{;cd}{} g^{cd},
\end{equation}
and
\begin{equation}
{\hat T}_i^{~k} = \Omega^{(-)4} T_i^{~k}.
\end{equation}
Here $;a$ is covariant derivative with respect Levi-Civita connection of the metric
in the initial gauge $g_{ab}(x)$.

A spacetime is {\it conformally flat} if there exist holonomic
coordinates $(x^0 =t,~x^1 = x,~ x^2 =y,~x^3 =z)$ in which its line element $ds^2$ has the
form
\begin{eqnarray}
ds^2 &=& \Omega^2(x^0,x^1,x^2,x^3) \bigl(dx^{0^2} - dx^{1^2} -
dx^{2^2}-dx^{3^2}\bigr)\nonumber\\
&\equiv& \Omega^2(x^0,x^1,x^2,x^3) \eta_{ik} dx^i dx^k.
\end{eqnarray}
The
\begin{equation}
\eta_{ik} dx^i dx^k = dx^{0^2} - dx^{1^2} - dx^{2^2} - dx^{3^2}
\end{equation}
means the line element of the empty, flat Minkowski spacetime in
inertial coordinates.

The Theorem is true:

Necessary and sufficient condition of the conformal flateness of
the 4-dimensional (or higher, $n>4$ dimensional) spacetime is vanishing
its {\it Weyl tensor} $C_{abcd}$, where
\begin{eqnarray}
C_{abcd}&:=& R_{abcd} + {R(g_{ac} g_{bd} - g_{ad}
g_{bc})\over(n-1)(n-2)} -\nonumber\\
&-& {\bigl(g_{ac} R_{bd} - g_{bc} R_{ad} + g_{bd} R_{ac} - g_{ad}
R_{bc}\bigr)\over (n-2)}.
\end{eqnarray}

In the above formula $R_{abcd}$ are components of the Riemann
tensor, $R_{ab}$ denote Ricci tensor components and $R$  means
Riemannian curvature scalar.

In the framework of general relativity Weyl's tensor $C_{abcd}$
describes free gravitational field (tidal forces).

An example of the conformally flat spacetimes give Friedman universes.

\section{Conformal transformation as Creator of the Friedman
universes}

We have under conformal rescaling of the metric (12) if we use the
formulas (14)-(15)
\begin{eqnarray}
{\hat G}_b^d &=& {\hat R}_b^d - {1\over 2}\delta_b^d {\hat R} =
\Omega^{(-)2} G_b^d \nonumber\\
&+& {2\over\Omega}\bigl(\Omega^{(-)1}\bigr)_{;bc} g^{dc}+ {3\over
\Omega^3}\delta^d_b \Omega_{;ce} g^{ce}\nonumber\\
&-&{1\over 2\Omega^4}\bigl(\Omega^2\bigr)_{;ac}{} g^{ac}
\delta_b^d;
\end{eqnarray}
\begin{equation}
{\hat T}_b^d= \Omega^{(-)4} T_b^d.
\end{equation}

By using Einstein equations in {\it old gauge} $g_{ik}(x)$
\begin{equation}
G_b^d = \beta T_b^d
\end{equation}
one can combine (19)-(20) to the form
\begin{equation}
{\hat G}_b^d = \beta\Omega^2 {\hat T}_b^d +  \beta {\widetilde
T}_b^d,
\end{equation}

where
\begin{eqnarray}
{\widetilde T}_b^d &:=&
{1\over\beta}\bigl[{2\over\Omega}\bigl(\Omega^{(-)1}\bigr)_{;bc}{}g^{dc}\nonumber\\
&+& {\delta_b^d\over\Omega^3}\bigl(3\Omega_{;ce}{}g^{ce} -
{\Omega^2_{~;ac}\over 2\Omega}{}g^{ac}\bigr)\bigr].
\end{eqnarray}

(22) gives Einstein equations in {\it new gauge} ${\hat g}_{ik}(x)$.

The tensor ${\widetilde T}_b^d(x)$ is the energy-momentum tensor of this matter
{\it which was created} by conformal rescaling of the initial metric
$g_{ik}(x)$ while the tensor  ${\hat T}_b^d(x)$ is transformed , following (20),
the matter tensor $T_b^d(x)$ which have already existed in the {\it old gauge} $g_{ik}(x)$.

One can rewrite (22) to the form
\begin{equation}
{\hat G}_b^d = \beta {\bar T}_b^d,
\end{equation}
where
\begin{equation}
{\bar T}_b^d := \Omega^2{\hat T}_b^d + {\widetilde T}_b^d.
\end{equation}

Of course, the {\it total matter} tensor (25) is covariantly
conserved.

 Friedman universes are conformally flat. So, we can take in the
 case as ``initial conditions''
 \begin{equation}
 g_{ik}(x)= \eta_{ik},~~G_b^d = 0, ~~T_b^d = 0\longrightarrow
 {\hat T}_b^d(x) = 0,
 \end{equation}
 i.e., we can take empty Minkowskian spacetime as initial
 spacetime.
 Doing so, one can get the metric tensor of a Friedman universe
 in the form
 \begin{equation}
 {\hat g}_{ik}(x) =\Omega^2(x) \eta_{ik},
 \end{equation}
 where conformal factor $\Omega(x)$ depends on Friedman
 universe.

  Thus, metric ${\hat g}_{ik}(x)$ of a Friedman universe, i.e.,
  {\it whole geometry} of a Friedman universe can be obtained from empty
  Minkowskian spacetime by a suitable conformal rescaling of the
  Minkowskian metric. Material content of this universe
  can be easily obtained from Einstein equations
  \begin{equation}
  {\widetilde T}_b^d := {1\over\beta} {\hat G}_b^d,
  \end{equation}
  where ${\hat G}_b^d(x)$ is Einstein tensor calculated from ${\hat
  g}_{ik}(x)$ or , immediately, from equations (23).

As an example we will consider a flat Friedman universe.

In this case
\begin{equation}
{\hat g}_{ik}(x) = \Omega^2 (\tau) \eta_{ik}=\Omega^2(\tau)
(d\tau^2 - dx^2 - dy^2 - dz^2)
\end{equation}
with $\Omega(\tau) \equiv R(\tau)$. $\tau$ is here the so-called
{\it conformal time} \cite{Gar2}.

After a simple but tedious calculation one gets from (28) [or from (23)] that
\begin{eqnarray}
{\widetilde T}_0^{~0}& =& {3R^{\prime}\over\beta R^4}~~
(=\rho)\nonumber\\
{\widetilde T}_1^{~1}& =&{\widetilde T}_2^{~2} = {\widetilde
T}_3^{~3} = {1\over\beta R^3}\bigl(2 R^{\prime\prime} -
{R^{\prime^2}\over R}\bigr)~~ (= -p).
\end{eqnarray}
Here prime denotes derivation with respect conformal time $\tau$.

Other components of the energy-momentum tensor ${\widetilde T}_b^{~a}$
of the matter created by conformal rescaling (29) of the
Minkowskian metric are vanishing.

For the flat dust Friedman universe we obtain
\begin{equation}
ds^2 = R^2(\tau)\bigl(d\tau^2 -dx^2 - dy^2 -dz^2\bigr),
\end{equation}
where
\begin{equation}
R(\tau) = {A^3\over 9}\tau^2, ~~A =\bigl(6\pi\rho R^3\bigr)^{1/3} =
const.
\end{equation}

From that one gets
\begin{equation}
R^{\prime} = {2A^3\tau\over 9},~~R^{\prime\prime} ={2A^3\over 9},
R^{\prime\prime\prime} = 0,
\end{equation}
and higher derivatives also vanish.

In consequence, the material content of the universe following
(30)  reads
\begin{equation}
{\widetilde T}_0^{~0} = \rho = {972\over\beta A^6\tau^6}.
\end{equation}
The other components of the tensor ${\widetilde T}_a^{~b}$ are
vanishing, i.e., $p=0$ and stresses vanish (as it should be in the
case).

Thus, we have correctly created flat, dust Friedman universe
from empty Minkowskian spacetime by using the conformal transformation
(31)-(32).

\section{Conclusion}
As we could see, Friedman universes {\it can be created} by a suitable
conformal rescaling of the flat Minkowskian metric,i.e., these
universes {\it can be created} from empty, flat Minkowskian spacetime by conformal
transformations.

Therefore, we needn't any ``quantum gravity'' in order to explain origin of the Friedman
universes: classical conformal transformations are sufficient.

The analogical statement is, of course, correct for any other conformally
flat spacetime.
\vspace{0.3cm}
\begin{center}
{\bf Acknowledgments}
\end{center}

Author would like to thank Professor Julian \L awrynowicz for
possibility  to deliver lecture at {\bf Hypercomplex Seminar
2012} in B\c edlewo, Poland, July 2012, and Mathematical Institute of the
University of Szczecin for financial support (Grant 503-4000-230351).

\newpage
\begin{center}
{\bf Interesuj\c aca w\l asno\'s\'c modeli kosmologicznych Friedmana}
\end{center}

\vspace{0.3cm}
\begin{center}
Janusz Garecki
\end{center}

\begin{center}
Instytut Matematyki Uniwersytetu Szczeci\'nskiego
\end{center}

\begin{center}
 Streszczenie
\end{center}
W tej pracy pokazano, \.ze modele kosmologiczne Friedmana, kt\'ore
s\c a  podstaw\c a  wsp\'o\l czesnej kosmologii, mo\.zna wykreowa\'c z
pustej czasoprzestrzeni Minkowskiego przy pomocy odpowiedniej
transformacji konforemnej.

\end{document}